\documentclass[aps,prl,reprint,amsmath,amssymb,superscriptaddress,floatfix]{revtex4-1}
\usepackage{graphicx}
\usepackage[T1]{fontenc}
\usepackage[utf8]{inputenc}
\usepackage{hyperref}
\usepackage[version=4]{mhchem}
\usepackage{gensymb}
\usepackage{siunitx}
\usepackage{letltxmacro}
\usepackage{pdfpages}
\usepackage{pgffor}

\makeatletter
\AtBeginDocument{\let\LS@rot\@undefined}
\makeatother

\renewcommand{\S}{\vec{S}}
\renewcommand{\vec}[1]{\mathbf{#1}}
\LetLtxMacro{\originaleqref}{\eqref}
\renewcommand{\eqref}{Eq.~\originaleqref}
\newcommand{\figref}[1]{Fig. \ref{#1}}
\newcommand{\figsref}[1]{Figs. \ref{#1}}

\newcommand{\ADDSNU}{\affiliation{Department of Physics and Astronomy, Seoul National University, Seoul 08826, Republic of Korea}}
\newcommand{\ADDIBS}{\affiliation{Center for Correlated Electron Systems, Institute for Basic Science (IBS), Seoul 08826, Republic of Korea}}
\newcommand{\ADDUOS}{\affiliation{Department of Physics, University of Seoul, Seoul 02504, Republic of Korea}}
\newcommand{\ADDUOSsecond}{\affiliation{Natural Science Research Institute, University of Seoul, Seoul 02504, Republic of Korea}}

\begin{document}

\title{Magnon topology and thermal Hall effect in trimerized triangular lattice antiferromagnet}

\author{Kyung-Su Kim}
\ADDSNU
\ADDIBS

\author{Ki Hoon Lee}
\email{kihoonlee@snu.ac.kr}
\ADDSNU
\ADDIBS

\author{Suk Bum Chung}
\ADDUOS
\ADDUOSsecond

\author{Je-Geun Park}
\ADDSNU
\ADDIBS


\begin{abstract}
The non-trivial magnon band topology and its consequent responses have been extensively studied in two-dimensional magnetisms.
However, the triangular lattice antiferromagnet (TLAF), the best-known frustrated two-dimensional magnet, has received less attention than the closely related Kagome system, because of the spin-chirality cancellation in the umbrella ground state of the undistorted TLAF.
In this work, we study the band topology and the thermal Hall effect (THE) of the TLAF with (anti-)trimerization distortion under the external perpendicular magnetic field using the linearized spin wave theory.
We show that the spin-chirality cancellation is removed in such case, giving rise to the non-trivial magnon band topology and the finite THE.
Moreover, the magnon bands exhibit band topology transitions tuned by the magnetic field.
We demonstrate that such transitions are accompanied by the logarithmic divergence of the first derivative of the thermal Hall conductivity.
Finally, we examine the above consequences by calculating the THE in the hexagonal manganite \ce{YMnO3}, well known to have anti-trimerization.
\end{abstract}
\maketitle

\textit{Introduction}.--The band topology has been extensively studied in the last decade for various quasiparticle excitations \cite{RevModPhys.89.040502, hasan_colloquium_2010, qi_topological_2011, ando_topological_2013, bernevig_topological_2013, chiu_classification_2016, haldane_photonicTI_2008, wang_photonicTI_2008, mcclarty2017topological,susstrunk2015observation}, including magnon, the elementary excitation of a magnetically ordered system.
As the magnetic ordering breaks the time-reversal symmetry (TRS), it is natural to expect the magnon band structures analogous to that of the Weyl semi-metal and the anomalous quantum Hall insulator \cite{li2016weyl,hwang2017magnon, Weylmagnon_pyro, Weyl_honeycomb_ferro, owerre2016magnon, owerre2017topological,  chisnell2015topological, laurell2018magnon, kagome_ferro_Mook, zhang_magnonTI_2013}.
In collinear phases, many magnon models with topological band structures were proposed \cite{zyuzin2016magnon, cheng2016SNE, lee2018magnonic}; however, there are only a few studies in non-collinear magnetic phases \cite{laurell2018magnon, Weyl_stacked_kagome_antiferro, owerre2016magnon}.

The band topology is closely related to transverse transport, and the thermal Hall effect (THE) is the most distinctive response expected from a time-reversal symmetry (TRS) broken phase like magnetism.
But the TRS breaking is only a necessary condition for having a finite THE, and whether a system shows a finite THE or not is also determined by its crystal symmetry.
For example, in the collinear antiferromagnetic honeycomb lattice, while the spin Nernst effect can be non-zero, the THE is forbidden by the symmetry constraint that forces a zero Chern number \cite{zyuzin2016magnon, cheng2016SNE, lee2018magnonic}.
More generally, a co-planar magnetic system without the spin-orbit coupling (SOC) cannot host a finite THE due to the effective TRS, which is the combination of the ordinary time-reversal and the $\pi$ spin rotation around the axis normal to the spin plane \cite{suzuki2017cluster}. 
Hence, in the absence of SOC, a non-coplanar spin configuration is necessary for the finite THE.

The simplest and yet most studied two-dimensional model hosting a non-coplanar phase is a triangular lattice antiferromagnet (TLAF) under a perpendicular external magnetic field.
However, to our best knowledge, in spite of the non-coplanar spin ordering, little is known of the magnon band topology and transverse response of TLAF.
This is due to the chirality cancellation in an undistorted TLAF, prohibiting a finite THE.
This situation is very different from a Kagome lattice, another archetypical frustrated lattice, which has been widely studied in various contexts of band topology \cite{chisnell2015topological, laurell2018magnon, kagome_ferro_Mook,Weyl_stacked_kagome_antiferro}.

In this work, we study the magnon band topology and the THE of the (anti-)trimerized TLAF under the perpendicular magnetic field using the linearized spin wave theory.
We clarify that the effective $PT$ symmetry forces THE to vanish for the undistorted TLAF even under the perpendicular magnetic field.
However, the (anti-)trimerization distortion removes this symmetry, allowing a finite THE.
Our study shows how the band topology of the TLAF depends on the distortion strength and the magnetic field, with the gap closings at the band topology transitions.
Remarkably, these transitions are characterized by a logarithmic divergence in the first derivative of the thermal Hall conductivity.
Such singularity behavior may be experimentally observed at zero magnetic field, where we found the strongest divergence due both to a quadratic band crossing at $\Gamma$ and multiple linear band crossings.
Finally, as a real experimental system we estimate the size of magnon THE in \ce{YMnO3}, the material in which the anti-trimerization is already observed \cite{park2008multiferroic}.

\begin{figure}[t]
\centering
\includegraphics[width=0.95\columnwidth]{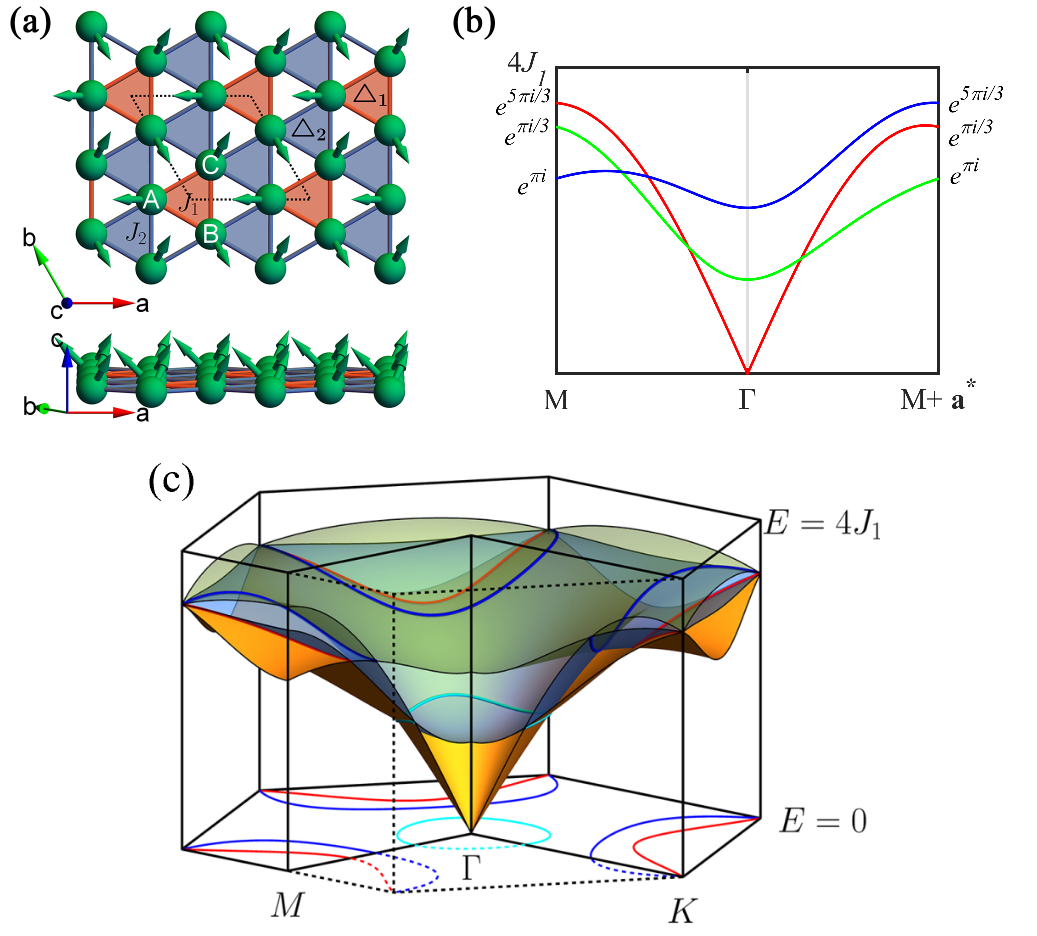}
\caption{(a) The umbrella phase of TLAF.
Red and blue equilateral triangles have different Heisenberg exchange coupling constants $J_1$ and $J_2$, respectively.
(b) The magnon band structure of the $J_1=J_2=1$ case with $h=1$ and $D^z=0.3$ along $M$ to $M+\vec{a}^*$.
Each band is classified according to the eigenvalues $\lambda e^{-i\vec{k}\cdot\vec{t}}$ of $\left\{g|\vec{t}\right\}$: $\lambda=1$ (red), $e^{2\pi i/3}$ (green) and $e^{-2\pi i/3}$ (blue).
Numbers on the left and right are $\lambda e^{-i\vec{k}\cdot\vec{t}}$ evaluated at the corresponding momentum.
(c) The magnon band structure in the whole momentum space.
The lines on the $E=0$ plane is the projection of nodal lines.}
\label{fig:ground_state_and_band}
\end{figure}

\textit{Model}.--In this work, we study the following anisotropic spin model on a triangular lattice with a magnetic field along the $z$-axis:
\begin{align}
{\cal H} =  &  J_{1} \sum_{\textrm{intra}} \S_{i} \cdot \S_{j}
              +J_{2} \sum_{\textrm{inter}} \S_{i} \cdot \S_{j}  
              +D^{z} \sum_{i}              (\textrm{S}_{i}^{z})^{2} \nonumber \\ 
            & -h     \sum_{i}             \textrm{S}_{i}^z,
\label{eq:trimerizedTLAF}
\end{align}
where $J_{1}>0$, $J_{2}>0$ and $D^{z}>0$.
Here $J_{1}$ and $J_{2}$ denote intra- and inter-trimer exchange constants as shown in \figref{fig:ground_state_and_band} (a), and $h=g\mu_{B}B$, where $g \approx 2$ is the g-factor and $\mu_{B}$ is the Bohr magneton.
    
First, we discuss the ground state of our model \eqref{eq:trimerizedTLAF}, where, we assume classical spins ($S\rightarrow\infty$).
The undistorted triangular lattice (i.e., $J_{1}=J_{2}$) has been studied extensively \cite{Starykh2011deformedTLAF,Yamamoto2014XXZTLAF,Eggert2015XXZTLAF}.
In such a case, for $D^{z}=0$, we have a 3-sublattice ground state structure subject to the constraint $\S_{A}+\S_{B}+\S_{C}=\vec{M}_{\triangle} = \hat{\bf z} h/3J,$ where $A,B$ and $C$ are the indices of the spins making a triangle as depicted in \figref{fig:ground_state_and_band} (a) and $M_{\Delta}$ is the sum of the spins.
This constraint fixes only three out of six free parameters (two for each spin sublattice) so that the classical ground state manifold is highly degenerate.
Adding a single-ion easy-plane anisotropy (or a two-ion anisotropy as in a XXZ model) lifts this classical accidental degeneracy, selecting an umbrella structure as the unique ground state \cite{Yamamoto2014XXZTLAF}.
A similar argument can be made for the case $J_1 \neq J_2$ by rearranging the Hamiltonian of \eqref{eq:trimerizedTLAF} as
\begin{align}
  \cal{H} =
   & 
        J_{1} \sum_{\triangle \in \triangle_{1}} 
               \left[\vec{M}_{\triangle} - \hat{\bf z} \frac{h}{3J_{\textrm{eff}}}\right]^{2}
      + J_{2} \sum_{\triangle \in \triangle_{2}} 
               \left[\vec{M}_{\Delta} - \hat{\bf z} \frac{h}{3J_{\textrm{eff}}} \right]^{2} \nonumber\\
   &  + D^{z} \sum_{i}{(\textrm{S}_{i}^z)^{2}} + (\textrm{const}),
  \label{rewrittenTLAF}
\end{align}
where $3J_{\textrm{eff}}=J_{1}+2J_{2}$, and $\triangle_1$ and $\triangle_2$ are both the sets of equilateral triangles but with different side lengths as in \figref{fig:ground_state_and_band} (a).
It can be also readily shown that in the absence of an easy-plane anisotropy we have the same 3-sublattice structure subject to the constraint
\begin{equation}
\S_{A}+\S_{B}+\S_{C}=\hat{\bf z}\frac{h}{3J_{\textrm{eff}}}.
\label{constraint}
\end{equation}
Now the easy-plane anisotropy selects the umbrella ground state as in the $J_1=J_2$ case.
And by a suitable parametrization of the spins in sublattices, i.e. $\S_{\alpha}=(\sin\theta\cos\phi_{\alpha}, \sin\theta\sin\phi_{\alpha}, \cos\theta)$ with $\phi_{\alpha}$'s making $120\degree$ to each other, we find the tilting angle $\theta = \cos^{-1}(h/h_{\textrm{c}})$, where $h_{\textrm{c}}=(9J_{\textrm{eff}}+2D^{z})S$.
Hence, our model \eqref{eq:trimerizedTLAF} has a simple ground state phase diagram with the umbrella structure below the saturation field $h_{c}$ and the fully polarized phase above.
We note that even though the quantum fluctuation favors competing coplanar phases 
over the umbrella phase, a sufficiently large easy-plane anisotropy and/or an antiferromagnetic interlayer coupling stabilizes the umbrella phase \cite{marmorini2016umbrella}.

\textit{Spin wave analysis}.--We perform the Holstein-Primakoff (HP) transformation on \eqref{eq:trimerizedTLAF} with the umbrella structure ground state: $\textrm{S}^n=S-a^{\dagger}a$, $\textrm{S}^+\simeq\sqrt{2S}a$ and $\textrm{S}^{-}\simeq\sqrt{2S}a^{\dagger}$, where $n$ is the local magnetization direction.
\begin{equation}
    \mathcal{H} = \frac{1}{2}\sum_{\alpha\beta\vec{k}} \psi^{\dagger}_{\alpha\vec{k}} H_{\alpha\beta}(\vec{k}) \psi_{\beta\vec{k}},
\end{equation}
where $\psi^{\dagger}_{\alpha\vec{k}} = \left[a^{\dagger}_{\alpha,\vec{k}}, a_{\alpha,-\vec{k}} \right]$ and $a_{\alpha,\vec{k}}$ is the HP boson operator of sublattice $\alpha= A,\ B,\ C$ and momentum $\vec{k}$.
The diagonalized form of the Hamiltonian is $\mathcal{H} = \frac{1}{2}
\sum_{\eta,\vec{k}}\left(E_{\vec{k}} \gamma^{\dagger}_{\eta\vec{k}} \gamma_{\eta \vec{k}} + E_{\eta,-\vec{k}} \gamma_{\eta,-\vec{k}}\gamma^{\dagger}_{\eta,-\vec{k}}\right)$, where $\sum_{\eta} \left[\gamma^{\dagger}_{\eta,\vec{k}},\gamma_{\eta,-\vec{k}} \right] T^{\dagger}_{\eta\alpha\vec{k}} = \psi_{\alpha\vec{k}}^{\dagger}$ and $T_{\vec{k}}$ is the para-unitary matrix (i.e., $T_{\vec{k}}^{\dagger} \sigma_3 T_{\vec{k}} = T_{\vec{k}} \sigma_3 T_{\vec{k}}^{\dagger} = \sigma_3$) diagonalizing $H({\vec{k}})$ \cite{colpa1978diagonalization}.
The Berry curvature is then defined as $\Omega_{n}^{z}(\vec{k}) = i \epsilon_{\mu\nu z} \left[\sigma_3 \partial_{k_{\mu}} T_{\vec{k}}^{\dagger} \sigma_3 \partial_{k_{\nu}} T_{\vec{k}}\right]_{nn}$. 
There are three magnon bands from three sublattices with a gapless linear Goldstone boson near $\Gamma$ from the breaking of U(1) spin-rotation symmetry around the $z$ axis \cite{SM}.
\begin{figure}[t]
\centering
\includegraphics[width=\columnwidth]{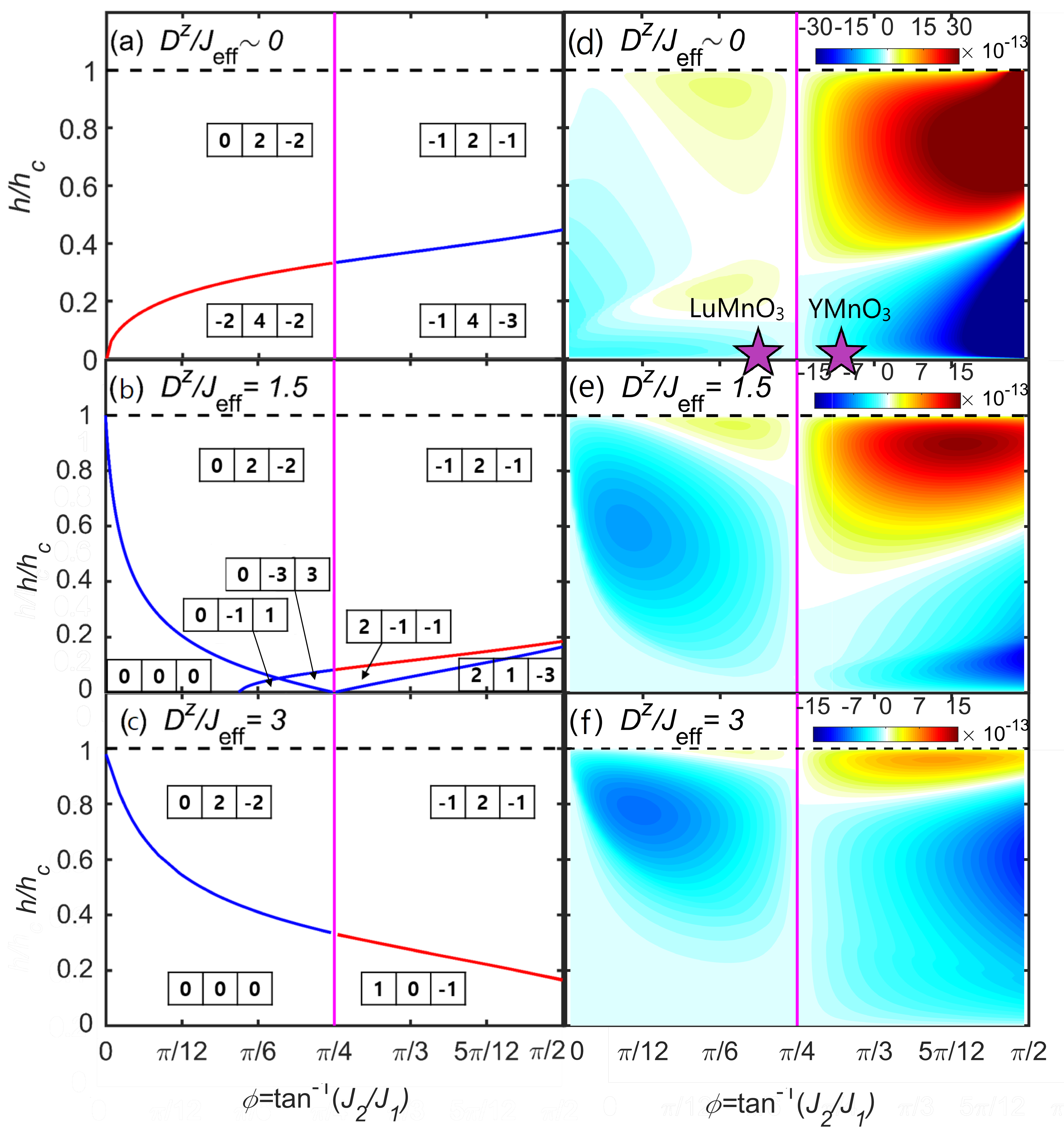}
\caption{(a)-(c) The band topology diagram for three different values of single-ion easy-plane anisotropy, where three Chern numbers, from the top band to the bottom bands, are denoted in the box. For (a) and (d), we assumed a small but finite easy-plane anisotropy in order to stabilize the umbrella ground state. (d) - (f) show the thermal Hall conductivity at $T=J_{\textrm{eff}}/k_B$. In (d) we display the expected trimerization magnitude $J_2/J_1\sim1.2$ and $J_2/J_1\sim0.8$ for \ce{YMnO3} and \ce{LuMnO3}, respectively.}
\label{fig:phase_diagram}
\end{figure}

Now in the undistorted case ($J_1=J_2$), a ``nonsymmorphic'' spin space group symmetry protects nodal lines and triple degenerate points at K points \cite{young2015dirac,young2017filling} (\figref{fig:ground_state_and_band} (b)-(c)).
To understand this, observe first that since the spin orderings of three sublattices in the umbrella state are related by the $120 \degree$ spin rotation around the $z$-axis, $\left\{g|\vec{t}\right\}=\left\{\exp{(2\pi i S^{z}/3)}|(2\vec{a}+\vec{b})/3 \right \}$ is the symmetry of the system, where $\vec{a}$ and $\vec{b}$ are the primitive lattice vectors (\figref{fig:ground_state_and_band} (a)).
Here $g$ leaves $\vec{k}$ invariant, and thus we can choose the Bloch states to be eigenstates of $\left\{g|\vec{t}\right\}$:  $\left\{g|\vec{t}\right\}\left|u_{\vec{k}}^{\lambda}\right\rangle=\lambda e^{-i\vec{k}\cdot \vec{t}}\left|u_{\vec{k}}^{\lambda}\right\rangle,$ where $\lambda=1,\, e^{\pm 2\pi i/3}$ since $\left\{g|\vec{t}\right\}^{3}\left|u_{\vec{k}}^{\lambda}\right\rangle=\left\{\hat{1}|\vec{2\vec{a}+\vec{b}}\right\}\left|u_{\vec{k}}^{\lambda}\right\rangle=e^{-3i\vec{k}\cdot \vec{t}}\left|u_{\vec{k}}^{\lambda}\right\rangle$.
Now, for $\vec{k} \rightarrow \vec{k}+\vec{a}^*$ ($\vec{k}+\vec{b}^*$), where $\vec{a}^*$ and $\vec{b}^*$ are the reciprocal primitive vectors corresponding to $\vec{a}$ and $\vec{b}$, we have $\lambda \rightarrow \lambda e^{2\pi i/3}$ ($\lambda e^{-2\pi i/3}$), alternating among the three eigenstates (\figref{fig:ground_state_and_band} (b)).
Thus, in general the three magnon branches meet even number of times (at least twice) as they cross BZ.
In \figref{fig:ground_state_and_band} (b), we show the case for $M \rightarrow M+\vec{a}^*$ where there are four such crossings.
This story holds for any $\vec{k} \rightarrow \vec{k}+\vec{a}^*$, leading to the nodal line structures as shown in \figref{fig:ground_state_and_band} (c). 
In addition, at two K points we have additional $C_{3z}$ symmetry, relating three eigenstates: $\left\{g|\vec{t}\right\} C_{3z}\left|u_{\vec{k}}^{\lambda}\right\rangle = C_{3z} \left\{g|C_{3z}^{-1}\vec{t}\right\} \left|u_{\vec{k}}^{\lambda}\right\rangle = \lambda e^{\pm 2\pi i /3} e^{-i\vec{k}\cdot \vec{t}}C_{3z}\left|u_{\vec{k}}^{\lambda}\right\rangle.$
Therefore, $\left\{g|\vec{t}\right\}$ and $C_{3z}$ protect the three-fold degeneracy at two K points.

All the nodal lines and triple degenerate points are gapped in the presence of the trimerization distortion in \eqref{eq:trimerizedTLAF}, as the spin nonsymmorphic symmetry is no longer present, generating the Berry curvature $\Omega_n^z(\vec{k})$ near these gaps.
Since the three bands are now gapped, the Chern number for individual band can be defined.

\textit{Band topology and thermal Hall effect}.--
In \figref{fig:phase_diagram}, we show Chern numbers $C_n=\frac{1}{2\pi}\int_{BZ}\Omega_n^z(\vec{k})d^2k $ for each band with the band topology transition lines and the intrinsic thermal Hall conductivity in the $h - J_{2}/J_{1}$ space for several values of $D^{z}$.
Here we used the parametrization $J_{1}=\cos\phi$ and $J_{2}=\sin\phi$.
The thermal Hall conductivity at temperature $T=J_{\textrm{eff}}/k_{B}$ is calculated using the following formula \cite{matsumoto2011rotational, matsumoto2011theoretical}: 
\begin{equation}
\kappa_{xy}=\frac{k_{B}^2T}{(2\pi)^2\hbar}\sum_{n}\int_{\textrm{BZ}}c_{2}(\rho_{\textrm{n}, \vec{k}})\Omega^{z}_{n}(\vec{k})d^{2}k,
\label{thermalHall}
\end{equation}
where $\Omega_n^z(\vec{k})$ is the Berry curvature of the $n$-th band at momentum $\vec{k}$.
Here, the $c_{2}$ function is given by $c_{2}(\rho)=(1+\rho)(\log\frac{1+\rho}{\rho})^{2}-(\log{\rho})^{2}-2\textrm{Li}_{2}(-\rho)$ with $\textrm{Li}_{2}(z)$ the polylogarithm function and $\rho_{\textrm{n}, \vec{k}}=1/(\exp{(\epsilon_{\textrm{n}, \vec{k}}/k_{B}T)}-1)$, where $\epsilon_{\textrm{n}, \vec{k}}$ is the energy of the $n$-th band at momentum $\vec{k}$ and $\rho$ is the Bose distribution function.
$T$, $\epsilon_{n}(\vec{k})$ and thus $\kappa_{xy}$ are normalized in units of $S\sqrt{J_{1}^{2}+J_{2}^{2}}$.

Let us make a general remark on the condition for a finite Chern number and THE in magnetic systems.
First, note that even though the TRS is broken, the magnetic systems with the coplanar spin ordering possess effective TRS in the absence of SOC, forbidding finite Hall responses \cite{suzuki2017cluster}.
More explicitly, the time reversal followed by $180\degree$ spin rotation around the axis normal to the plane, $\Tilde{T}=\exp({-i\pi S^{z}})T$, is the symmetry of the system in such a case.
This symmetry imposes a constraint on the Berry curvature, $\Omega^{z}_{n}(\vec{k})=-\Omega_{n}^z(-\vec{k})$, enforcing both the band Chern number and the thermal Hall conductivity to be zero.
However, if the system possesses a non-coplanar spin configuration with nonzero chirality $\chi = \S_{A} \cdot \S_{B} \times \S_{C}$, then the effective time reversal symmetry is broken and we can expect a finite band Chern number and THE.

However, following the line (magenta) for an undistorted triangular lattice ($J_{1}=J_{2}$) in \figref{fig:phase_diagram}, THE is zero even when finite magnetic field is applied.
It is because of the effective $PT$ symmetry $\tilde{I}=\exp({-i\pi S^{y}})PT$, where inversion center is at the middle of B and C in \figref{fig:ground_state_and_band}.(a).
In this case, we have $\Omega^z_n(\vec{k})=-\Omega^z_n(\vec{k})=0$, also forbidding a finite Chern number and THE.
This situation can also be understood heuristically in terms of spin chirality $\chi$: because $\chi$ has the opposite sign for the neighboring triangles, it cancels out and magnon feels no gauge field.
Note that the constraint from the effective $PT$ symmetry applies to the charge Hall effect in itinerant magnetic systems as well.
In the presence of the (anti-)trimerization ($J_{1} \neq J_{2}$), the effective $PT$ symmetry is absent.
Now since there is no symmetry to enforce the Berry curvature to vanish, we expect a finite magnon THE.

We find that the magnon band structure exhibits a rich band topology diagram in the $h-J_2/J_1$ space.
Since the three bands are separated from one another away from the $J_1=J_2$ lines (magenta) and below the saturation field (dashed line) in the parameter space, the Chern numbers are well defined for the top, middle and bottom bands, as denoted in \figsref{fig:phase_diagram} (a)-(c). 
On the red lines, there is an accidental gap closing between the top and middle bands, while on the blue lines, between the middle and bottom bands.
We further find that all the degeneracies occur either at K points or on the $\Gamma$-M segment.
Two accidental gap closings appear at the K point as the two K points are related by $M_{\vec{y}}T$ while three accidental gap closings appear for the $\Gamma$-M case as three $\Gamma$-M lines are related by $C_{3z}$.
Hence, the topological band transition with the gap closings at K changes the Chern number by $2$ and the one with the gap closings at $\Gamma$-M changes it by $3$. 

\begin{figure}[t]
\centering
\includegraphics[width=\columnwidth]{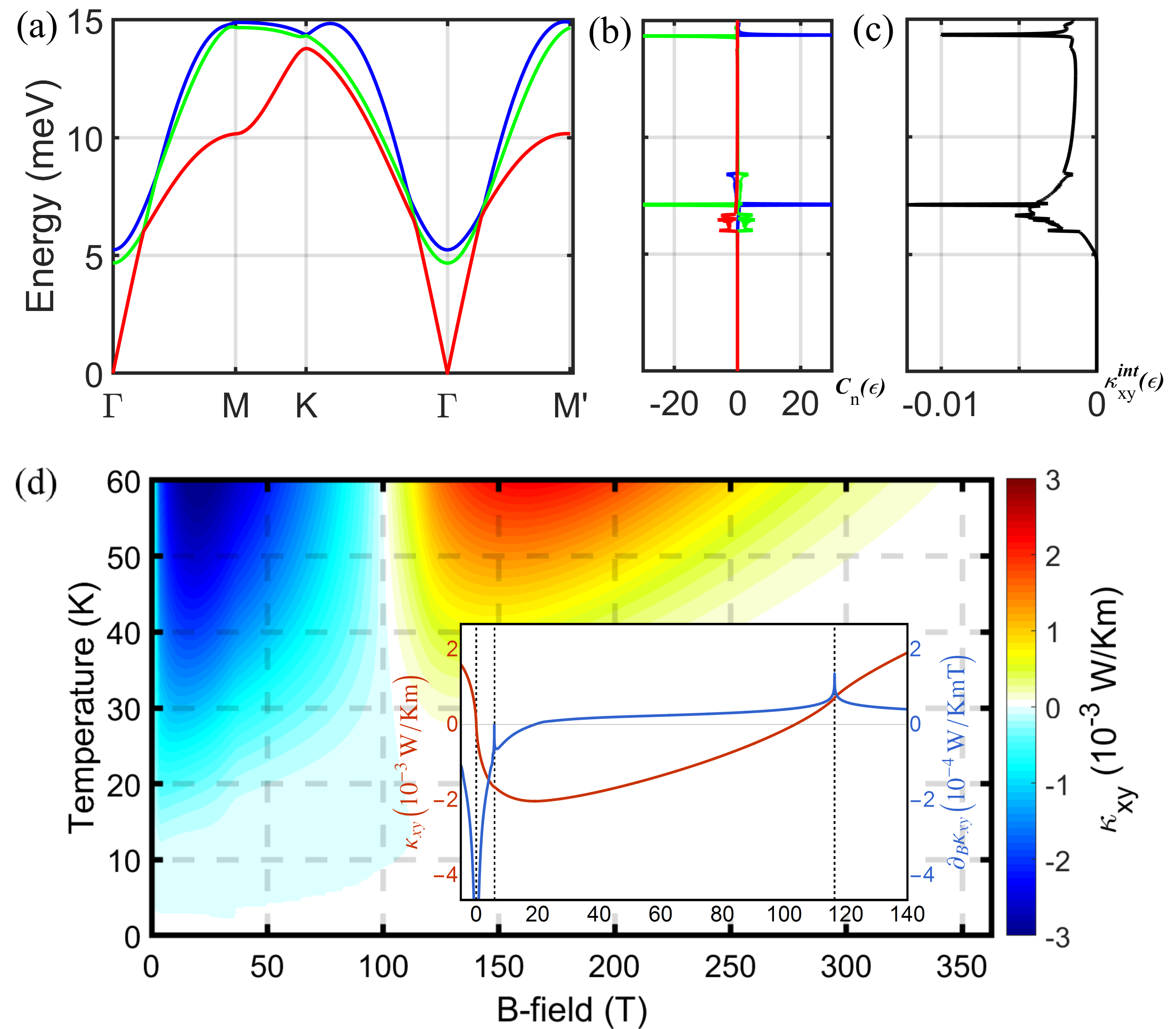}
\caption{(a) Magnon band structure of \ce{YMnO3} with
$B=\SI{5}{\tesla}$.
(b) Energy-resolved Chern number. (c) Integrated thermal Hall conductivity $\kappa_{xy}^{\textrm{int}}(\epsilon)$ as defined in the text at $T=\SI{40}{\K}$. The thermal Hall conductivity is $\kappa_{xy}=\SI{-1.53e-3}{\watt/\kelvin\meter}$. (d) Temperature and magnetic field dependence of calculated thermal Hall conductivity of \ce{YMnO3}. The inset is $\kappa_{xy}$ and $\partial_B \kappa_{xy}$ at \SI{40}{\kelvin}, where the singularities appear at the band topology transition points (dotted lines): $B=0,$ $5.9$ and $\SI{116.3}{\tesla}$. The band Chern numbers, from the top to the bottom band, are $[-2,-1,3]$ $(B<\SI{0}{\tesla})$, $[2,1,-3]$ $(\SI{0}{\tesla}<B<\SI{5.9}{\tesla})$, $[-1,4,-3]$ $(\SI{5.9}{\tesla}<B<\SI{116.3}{\tesla})$, and $[-1,2,-1]$ $(\SI{116.3}{\tesla}<B<\SI{363}{\tesla})$, respectively.
At $B=0$, $\kappa_{xy}$ and the band Chern numbers reverse their sign, because $B<0$ and $B>0$ sectors are related by $C_{2x}$.}
\label{fig:YMnO3}
\end{figure}

In \figsref{fig:phase_diagram} (d)-(f) we show the intrinsic contribution to the magnon thermal Hall conductivity.
It has intriguing behavior, yet at first sight reveals no apparent relation to the band topology diagram of \figsref{fig:phase_diagram} (a)-(c).
Interestingly, there was a numerical observation of the singular behavior of $\kappa_{xy}$ at the phonon band topology transition point \cite{topological_phonon}; however, the appropriate explanation was not provided.
Here, we found that the band topology transition of a free bosonic system manifests itself as the logarithmic divergence in the first derivative of $\kappa_{xy}$ at the transition point both for the linear and higher-order band crossing. 
To see it clearly, let us assume without loss of generality that the transition occurs at $p=0$, where $p$ could be any parameter inducing a band topology transition (e.g., external magnetic field) with gap closing between two bands at $\vec{k}=0$ and $E= \mathcal{E}_{0}$, leading to a Weyl point in the $\tilde{\vec{k}}=(k_x,k_y,p)$ space.
In the case of the isotropic single Weyl point, i.e. $H(\tilde{\vec{k}})= \mathcal{E}_{0}-\tilde{\vec{k}}\cdot\vec{\sigma}$, the singular contribution of the THE is $\tilde{\Omega}_{n,\vec{k}}^pc_2(\rho_{n,\vec{k}})=\pm \frac{p}{2(k^2+p^2)^{3/2}} c_2 (\rho(\mathcal{E}_0 \pm \epsilon_{\vec{k}}^p))$ for the upper and lower bands, respectively, where $\epsilon^p_{\vec{k}}=\sqrt{k^2+p^2}$.
Now, since $c_2(\rho(\mathcal{E}_0 + \epsilon^p_{\vec{k}}))-c_2(\rho(\mathcal{E}_0 - \epsilon^p_{\vec{k}}))\propto\epsilon^p_{\vec{k}}$ for small $k$ and $p$, one immediately notices the logarithmic divergence of the first derivative of $\tilde{\kappa}_{xy}$ at the transition point $p=0$:
\begin{equation}
    \frac{\partial}{\partial p}\tilde{\kappa}^p_{xy} \propto \frac{\partial}{\partial p} \int_{k<k_c} d^2 k\Omega_{n,\vec{k}} \epsilon^p_{\vec{k}} \propto \log \left| p \right| + \cdots.
    \label{eq:logdivergence}
\end{equation}
This result can be easily generalized to include multiple gap closings and the anisotropy.
In the case of the multi-Weyl point, $H(\tilde{\vec{k}})= \mathcal{E}_{0}-(k^n\cos(n\phi),k^n\sin(n\phi) ,p)\cdot\vec{\sigma}$, where $\tan\phi=k_y/k_x$, we have $\tilde{\Omega}_{n,\vec{k}} \epsilon^p_{\vec{k}}=\frac{pn^2k^{2(n-1)}}{k^{2n}+p^2}$, leading to the same logarithmic singularity but with a higher-order band crossing at the transition point.
We note that our model exhibits both accidental linear band crossings and essential quadratic band crossings; the latter occurs at $\Gamma$ when $h=0$ due to the combination of $C_3$ point symmetry group and $\tilde{T}$ \cite{SM}.
We corroborate the above results by the numerical calculation of the THE for \ce{YMnO3} (the inset of \figref{fig:YMnO3} (d)), which we will discuss below in more detail.

\textit{Thermal Hall Effect in} \ce{YMnO3}.--
As a real experimental example, we consider \ce{YMnO3}, in which $\ce{Mn^{3+}}$ ions with $S=2$ form a quasi-2D TLAF with the interlayer distance of $c/2= \SI{5.7}{\angstrom}$.
The strong spin-lattice coupling was previously reported in this multiferroic material, where the anti-trimerization distortion of the lattice occurs at $T_{N}=\SI{75}{\kelvin}$ \cite{park2008multiferroic}. 
The magnetic structure and the origin of multiferroicity of the compound have long been the subject of debate.
In particular, the lattice distortion was at first thought to lead to a large modulation in the exchange constant $J$ \cite{oh2013magnon, lewtas2010magnetic}, but a more recent study shows a smaller (yet still significant) effect with $J_{2}/J_{1}\sim 1.2$  \cite{JSOh2016mpcoupling, Lepetit2013abinitio}. 
Here, we assume the realistic parameters for the magnetic Hamiltonian of \ce{YMnO3} and calculate the THE: $J_{1}=\SI{2}{\meV}$, $J_{2}=\SI{2.4}{\meV}$ and $D^{z}=\SI{0.3}{\meV}$.

First, we show the magnon band structure along with energy-resolved Chern number $C_{\textrm{n}}(\epsilon)=\frac{1}{2\pi}\int_{\textrm{BZ}}\delta(\epsilon_{\textrm{n}, \vec{k}}-\epsilon)\Omega^{z}_{n}(\vec{k})d^{2}k$ and the integrated thermal Hall conductivity $\kappa_{xy}^{\textrm{int}}(\epsilon)=\frac{k_{B}^2T}{(2\pi)^2\hbar}\sum_{n}\int_{\epsilon_{\textrm{n}, \vec{k}} < \epsilon}c_{2}(\rho_{\textrm{n}, \vec{k}})\Omega^{z}_{n}(\vec{k})d^{2}k$ 
for $B=h/g\mu_{B}=\SI{5}{\tesla}$ and $T=\SI{40}{\kelvin}$ (\figref{fig:YMnO3} (a)-(c)). 
Each band shows an interesting Berry curvature contribution to the magnon thermal Hall conductivity.
We next show the magnetic field and temperature dependence of $\kappa_{xy}$, plotted up to the saturation field $B_{c} \simeq \SI{363}{\tesla}$ and $T=\SI{60}{\kelvin} < T_{N}$ (\figref{fig:YMnO3}. (d)).
We observe that the THE is still large ($\kappa_{xy}\sim \SI{-e-3}{\watt/\kelvin\meter}$) even for small magnetic fields ($\sim \SI{10}{\tesla}$) at temperatures as low as \SI{30}{K}.
We emphasize that since the longitudinal thermal conductivity is measured to be $\kappa_{xx}\sim \SI{10}{\watt/\kelvin\meter}$ \cite{sharma2004thermal}, the Hall angle $\kappa_{xy}/\kappa_{xx}\sim10^{-4}$ of our result is in an observable range of experiments \cite{ideue2017giant}.
Furthermore, the THE shows the most profound signature of the singularity of $\kappa_{xy}$, as derived in \eqref{eq:logdivergence}, at zero magnetic field, where both the linear and quadratic crossings occur.
Hence, we expect that the consequence of the band topology transition can be measured by a careful experiment in \ce{YMnO3}.

Before concluding, we remark that because of the large spin-lattice coupling in \ce{YMnO3} \cite{JSOh2016mpcoupling}, the phonon contribution to $\kappa_{xy}$ may not be negligible.
In fact, even in the absence of trimerization, the effective $PT$ symmetry, $\tilde{I}=\exp({-i\pi S^{y}})PT$, is still broken in the material when we consider non-magnetic ions such as $\ce{O^{2-}}$, and so magneto-elastic excitation may contribute appreciably to $\kappa_{xy}$.
We leave this issue as the focus of our future study.

\textit{Conclusion}.--In conclusion, we considered the trimerization distortion and the magnetic field on TLAF, which
give rise to the non-trivial band topology and the finite THE.
This leads to a variety of topologically distinct band structures, in contrast to a rather simple undistorted case \cite{katsura2010theory, owerre_bilayerTLAF}.
As one crosses the band topology transition boundary, the first derivative of the thermal Hall conductivity shows a logarithmic divergence.
This establishes the clear relation between the bosonic band topology and the THE.
We finally propose the hexagonal manganite family \ce{RMnO3} with the P6$_3$cm space group as the candidate material to detect such effects.
\begin{acknowledgments}
{\it Acknowledgement}:
We thank Kisoo Park and Haleem Kim for helpful discussions.
The work was supported by the Institute for Basic Science in Korea (IBS-R009-G1) (K.-S.K., K.H.L. and J.-G.P.) and Basic Science Research Program through the National Research Foundation of Korea (NRF) funded by the Ministry of Education (2018R1D1A1B07045899) (S.B.C.).
\end{acknowledgments}

\bibliography{TLAF_Top_ref}

\clearpage


\section*{Supplementary material}

# Supplemental Material for "Magnon topology and thermal Hall effect in trimerized triangular lattice antiferromagnet"

## 1 Hamiltonian Representation

We present the representation of Holstein-Primarkoff (HP) Hamiltonian for Eq. (1) of the main text. First, in the local frame of magnetization HP transformation is, up to quadratic order,

$$\mathbf{S}^{\alpha}_{\text{local}} = \left[ S - a^{\dagger}_{\alpha} a_{\alpha}, \sqrt{S/2}(a_{\alpha} + a^{\dagger}_{\alpha}), -i\sqrt{S/2}(a_{\alpha} - a^{\dagger}_{\alpha}) \right]^{T}, \quad (1)$$

where $\alpha = 0, 1, 2$ is the sublattice index. By suitable rotation, $\mathbf{S}_{\text{local}}$ can be transformed to the global frame:

$$\mathbf{S}^{\alpha}_{\text{global}} = \begin{bmatrix} \cos\phi^{\alpha} & -\sin\phi^{\alpha} & 0 \\ \sin\phi^{\alpha} & \cos\phi^{\alpha} & 0 \\ 0 & 0 & 1 \end{bmatrix} \begin{bmatrix} \cos\theta & 0 & -\sin\theta \\ 0 & 1 & 0 \\ \sin\theta & 0 & \cos\theta \end{bmatrix} \mathbf{S}_{\text{local}}, \quad (2)$$

where $\theta = \sin^{-1}(h/h_c)$ is the tilting angle as defined in the text and $\phi^{\alpha}$'s are making $120°$. Then, we define the Fourier transformation, $\psi_{\mathbf{k}} = \left[a_{1,\mathbf{k}}, a_{2,\mathbf{k}}, a_{3,\mathbf{k}}, a^{\dagger}_{1,-\mathbf{k}}, a^{\dagger}_{2,-\mathbf{k}}, a^{\dagger}_{3,-\mathbf{k}}\right]^{T}$, of field operators $X_{\mathbf{R}_i} = \left[a_{1,\mathbf{R}_i}, a_{2,\mathbf{R}_i}, a_{3,\mathbf{R}_i}, a^{\dagger}_{1,\mathbf{R}_i}, a^{\dagger}_{2,\mathbf{R}_i}, a^{\dagger}_{3,\mathbf{R}_i}\right]^{T}$:

$$a_{\alpha, \mathbf{k}} = \frac{1}{\sqrt{N}} \sum_{\mathbf{R}_i + \tau_{\alpha}} e^{-i\mathbf{k} \cdot (\mathbf{R}_i + \tau_{\alpha})} a_{\alpha, \mathbf{R}_i}, \text{ or} \quad (3)$$

$$\psi_{\mathbf{k}} = \frac{1}{\sqrt{N}} \sum_{\mathbf{R}_i} e^{-i\mathbf{k} \cdot \mathbf{R}_i} \text{diag}\left(\left[e^{-i\mathbf{k} \cdot \tau_1}, e^{-i\mathbf{k} \cdot \tau_2}, e^{-i\mathbf{k} \cdot \tau_3}, e^{-i\mathbf{k} \cdot \tau_1}, e^{-i\mathbf{k} \cdot \tau_2}, e^{-i\mathbf{k} \cdot \tau_3}\right]\right) X_{\mathbf{R}_i} \equiv \frac{1}{\sqrt{N}} \sum_{\mathbf{R}_i} e^{-i\mathbf{k} \cdot \mathbf{R}_i} \Gamma(\mathbf{k}) X_{\mathbf{R}_i} \quad (4)$$

where $\mathbf{R}_i$s are lattice vectors and $\tau_{\alpha}$ is a sublattice vector, i.e. $\tau_1 = -\mathbf{a}/3$, $\tau_2 = -\mathbf{b}/3$ and $\tau_3 = (\mathbf{a} + \mathbf{b})/3$. The Hamiltonian can be written as:

$$\mathcal{H} = J_1 \sum_{\text{intra}} \mathbf{S}_i \cdot \mathbf{S}_j + J_2 \sum_{\text{inter}} \mathbf{S}_i \cdot \mathbf{S}_j + D^z \sum_i (S_i^z)^2 - h \sum_i S_i^z \quad (5)$$

$$= J_1 \sum_{\triangle \in \triangle_1} \left[\mathbf{M}_{\triangle} - \hat{\mathbf{z}} \frac{h}{3 J_{\text{eff}}}\right]^2 + J_2 \sum_{\triangle \in \triangle_2} \left[\mathbf{M}_{\triangle} - \hat{\mathbf{z}} \frac{h}{3 J_{\text{eff}}}\right]^2 + D^z \sum_i (S_i^z)^2 + (\text{const}) \quad (6)$$

$$= \frac{1}{2} \sum_{\mathbf{k}} \psi^{\dagger}_{\mathbf{k}} H(\mathbf{k}) \psi_{\mathbf{k}} + (\text{const}) \quad (7)$$

$$H(\mathbf{k}) = H_1(\mathbf{k}) + H_2(\mathbf{k}) + H_D(\mathbf{k}), \quad (8)$$

where $H_1(\mathbf{k})$, $H_2(\mathbf{k})$ and $H_D(\mathbf{k})$ are the Fourier components from $J_1$, $J_2$ and $D^z$ terms in (6). Now by applying (2), (4) to each terms in (6), we obtain

$$H_1(\mathbf{k}) = \frac{J_1 S}{2} \Gamma(\mathbf{k}) H_\theta \Gamma(\mathbf{k})^\dagger \tag{9}$$

$$H_2(\mathbf{k}) = \frac{J_2 S}{2} (U_2(\mathbf{k}) \Gamma(\mathbf{k}) H_\theta \Gamma(\mathbf{k})^\dagger U_2(\mathbf{k})^\dagger + U_3(\mathbf{k}) \Gamma(\mathbf{k}) H_\theta \Gamma(\mathbf{k})^\dagger U_3(\mathbf{k})^\dagger) \tag{10}$$

$$H_D(\mathbf{k}) = D^z S \cos^2\theta \, \Gamma(\mathbf{k}) \begin{bmatrix} \mathbf{1}_3 & -\mathbf{1}_3 \\ -\mathbf{1}_3 & \mathbf{1}_3 \end{bmatrix} \Gamma(\mathbf{k})^\dagger \tag{11}$$

$$H_\theta = \begin{bmatrix} T & S \\ S & T^* \end{bmatrix}, \quad T = \begin{bmatrix} 2 & t_\theta & t_\theta^* \\ t_\theta^* & 2 & t_\theta \\ t_\theta & t_\theta^* & 2 \end{bmatrix}, \quad S = \begin{bmatrix} 0 & s_\theta & s_\theta \\ s_\theta & 0 & s_\theta \\ s_\theta & s_\theta & 0 \end{bmatrix}, \tag{12}$$

$$t_\theta = i\sqrt{3}\sin\theta + \frac{3\cos^2\theta - 2}{2}, \quad s_\theta = -\frac{3}{2}\cos^2\theta \tag{13}$$

$$U_2(\mathbf{k}) = \mathrm{diag}\left([e^{-i\mathbf{k}\cdot\mathbf{a}}, 1, e^{i\mathbf{k}\cdot\mathbf{b}}, e^{-i\mathbf{k}\cdot\mathbf{a}}, 1, e^{i\mathbf{k}\cdot\mathbf{b}}]\right) \quad U_3(\mathbf{k}) = \mathrm{diag}\left([e^{-i\mathbf{k}\cdot\mathbf{a}}, e^{-i\mathbf{k}\cdot(\mathbf{a}+\mathbf{b})}, 1, e^{-i\mathbf{k}\cdot\mathbf{a}}, e^{-i\mathbf{k}\cdot(\mathbf{a}+\mathbf{b})}, 1]\right), \tag{14}$$

where $\mathbf{1}_3$ is the $3 \times 3$ identity matrix.

## 2  Symmetry analysis of the trimerized triangular lattice

| $J_1 \neq J_2$ & $h = 0$ (nonmagnetic) | $J_1 \neq J_2$ & $h = 0$ (coplanar) | $J_1 \neq J_2$ & $h \neq 0$ (non-coplanar) |
|---|---|---|
| 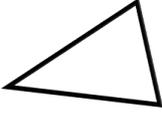 | 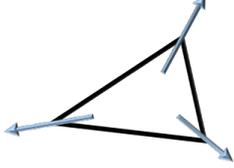 | 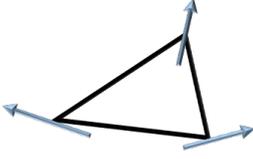 |
| $\{E, T\} \times SU(2) \times D_{3h}$ | $\{E, \tilde{M}_z\} \times \{E, \tilde{T}\} \times D_3$ | $\{E, \tilde{M}_z\} \times \{E, 2C_3, 3\sigma_v T\}$ |

Figure 1: The spin space group for trimerized triangular lattice in the case for nonmagnetic, 120° coplanar and non-coplanar phase, respectively. Here, $T$ and $M_z$ are the ordinary time-reversal and mirror operation, respective, and $\tilde{T} = T e^{-i\pi S_z}$ and $\tilde{M}_z = M_z e^{-i\pi S_z}$ are the effective time-reversal and mirror operation.

In this section, we present the full spin space group symmetry of the trimerized triangular lattice in the presence and absence of the perpendicular magnetic field. In the case for the nonmagnetic phase, we have, in addition to the lattice point symmetry group $D_{3h}$, the $SU(2)$ spin-rotation and the time-reversal symmetry. This symmetry is lowered in the magnetically ordered phase as shown in Fig. 1.

We now focus on the $h = 0$ case where the spin ordering is coplanar. First, there is a two-fold degeneracy at $\Gamma$ and $K$ due to a pair of complex-conjugate one-dimensional representations of $C_3$ related by $\tilde{T}$. More explicitly, acting $C_{3z}$ on $\mathbf{S}_{\mathrm{global}}^\alpha$ gives

$$C_{3z} \mathbf{S}_{\mathrm{global}}^\alpha C_{3z}^{-1} = R(C_{3z})^{-1} \mathbf{S}_{\mathrm{global}}^{\alpha+1 \ (\mathrm{mod}\ 3)}, \quad \text{or} \quad C_{3z} \mathbf{S}_{\mathrm{local}}^\alpha C_{3z}^{-1} = \mathbf{S}_{\mathrm{local}}^{\alpha+1 \ (\mathrm{mod}\ 3)}, \tag{15}$$

where $R(C_{3z})$ is the ordinary three-dimensional representation of $C_{3z}$ acting on the real space. We have chosen sublattice indices to be related by $C_{3z}$. We immediately see that $C_{3z}$ permutes three magnon operators: $C_{3z} a_\alpha C_{3z}^{-1} = a_{\alpha+1 \ (\mathrm{mod}\ 3)}$. We may write this as

$$C_{3z} \doteq \begin{bmatrix} 0 & 1 & 0 \\ 0 & 0 & 1 \\ 1 & 0 & 0 \end{bmatrix}. \tag{16}$$

2

The two eigenvalues $e^{\pm \frac{2\pi i}{3}}$ of this matrix form the pair of complex-conjugate representations of $C_3$ related by $\tilde{T}$, enforcing two-fold degeneracy at $\Gamma$ and $K$. Additionally, we show that the band touching at $\Gamma$ must be quadratic by constructing the effective two-band $\mathbf{k} \cdot \mathbf{p}$ Hamiltonian near $\Gamma$.

On the basis where $[1,0]^T$ and $[0,1]^T$ represent the eigenvectors corresponding to the eigenvalues $e^{\pm \frac{2\pi i}{3}}$ of $C_{3z}$, the symmetry allowed Hamiltonian up to quadratic in $k$ [1] is written as

$$H_{\text{eff}}(\mathbf{k}) = (vk_- + wk_+^2)\sigma_+ + (v^*k_+ + w^*k_-^2)\sigma_-, \tag{17}$$

where $k_\pm = k_x \pm ik_y$ and $\sigma_\pm = (\sigma_x \pm i\sigma_y)/2$. Now, the anti-unitary operator $\tilde{T}$ transforms the two eigenstates to each other and $\tilde{T}^2 = 1$ for bosons; therefore, $\tilde{T}$ is represented by $\sigma_x K$ where K is the complex-conjugation operator. It can be easily seen that $\tilde{T}$ leaves $H_{\text{eff}}(\mathbf{k})$ invariant: $\tilde{T}H_{\text{eff}}(\mathbf{k})\tilde{T}^{-1} = H_{\text{eff}}(\mathbf{k})$. But we also have $\tilde{T}H_{\text{eff}}(\mathbf{k})\tilde{T}^{-1} = H_{\text{eff}}(-\mathbf{k})$. Hence the k-odd terms are not allowed, leaving the quadratic term as the lowest allowed term of the effective Hamiltonian. The similar situation occurs for electronic bands in the presence of $C_{3z}$ and the ordinary time-reversal $T$, as demonstrated in [2]. Finally, there are six $C_{2x}$ symmetry allowed crossings on $\Gamma - K$ lines. Six such crossings are connected by $C_{3z}$ and $\tilde{T}$. These exhaust the symmetry enforced band crossing at $h = 0$. (See $h/h_c = 0$ case in Fig. 3.)

Now in the non-coplanar phase in the presence of the magnetic field, $\tilde{T}$ and $C_{2x}$ are no longer present, opening a gap at these crossings.

At a linear (quadratic) crossing, the Berry phase $\pm\pi$ ($\pm 2\pi$) is induced near the gap. (Fig. 3).

## 3 Band structure and Berry curvature $\phi = \pi/3$ & $D^z = 1.5J_{\text{eff}}$

At $h/h_c = 0$ and $h/h_c = 1$, where $h_c = (9J_{\text{eff}} + 2D^z)S$ is the saturation field, the magnetic ordering is co-planar and collinear, respectively, and thus we have zero Chern number and zero thermal Hall effect as explained in the main text. But in between, i.e. $0 < h/h_c < 1$, three bands may have nonzero Chern number. Below we show the band structures along high symmetry points and Berry curvature plots of the top, middle and bottom band on BZ for $\phi = \pi/3$ and $D^z = 1.5J_{\text{eff}}$ as increasing magnetic field $h$. Topological band structure transition occurs at $h/h_c = 0, 0.056$ (K point crossing) and $0.115$ ($\Gamma$-M crossing). (See Fig. 2 (b) of the main text.)

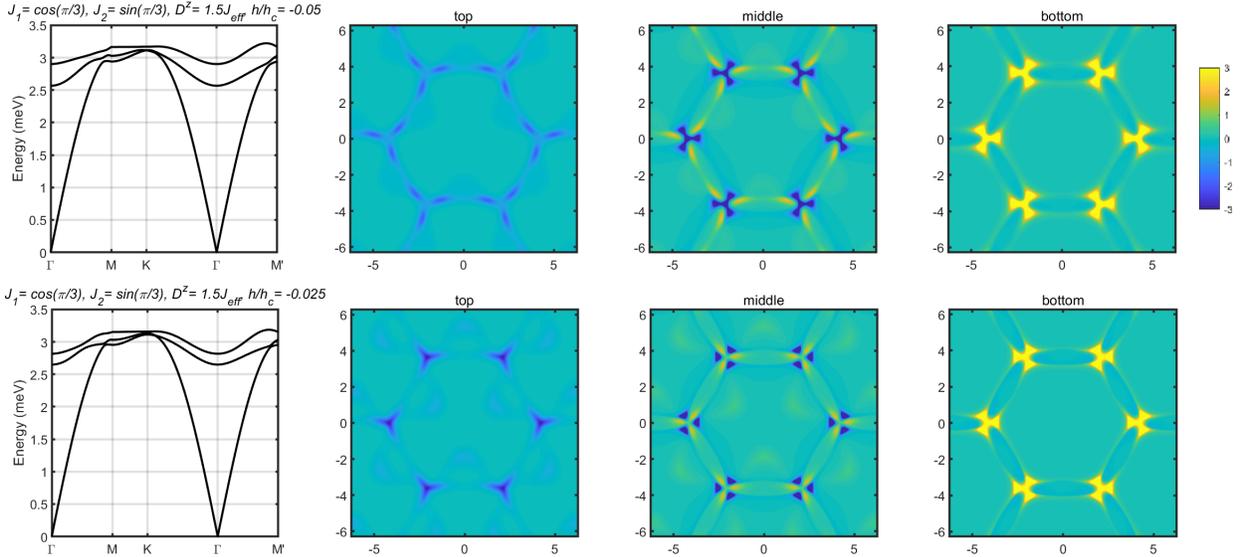

3

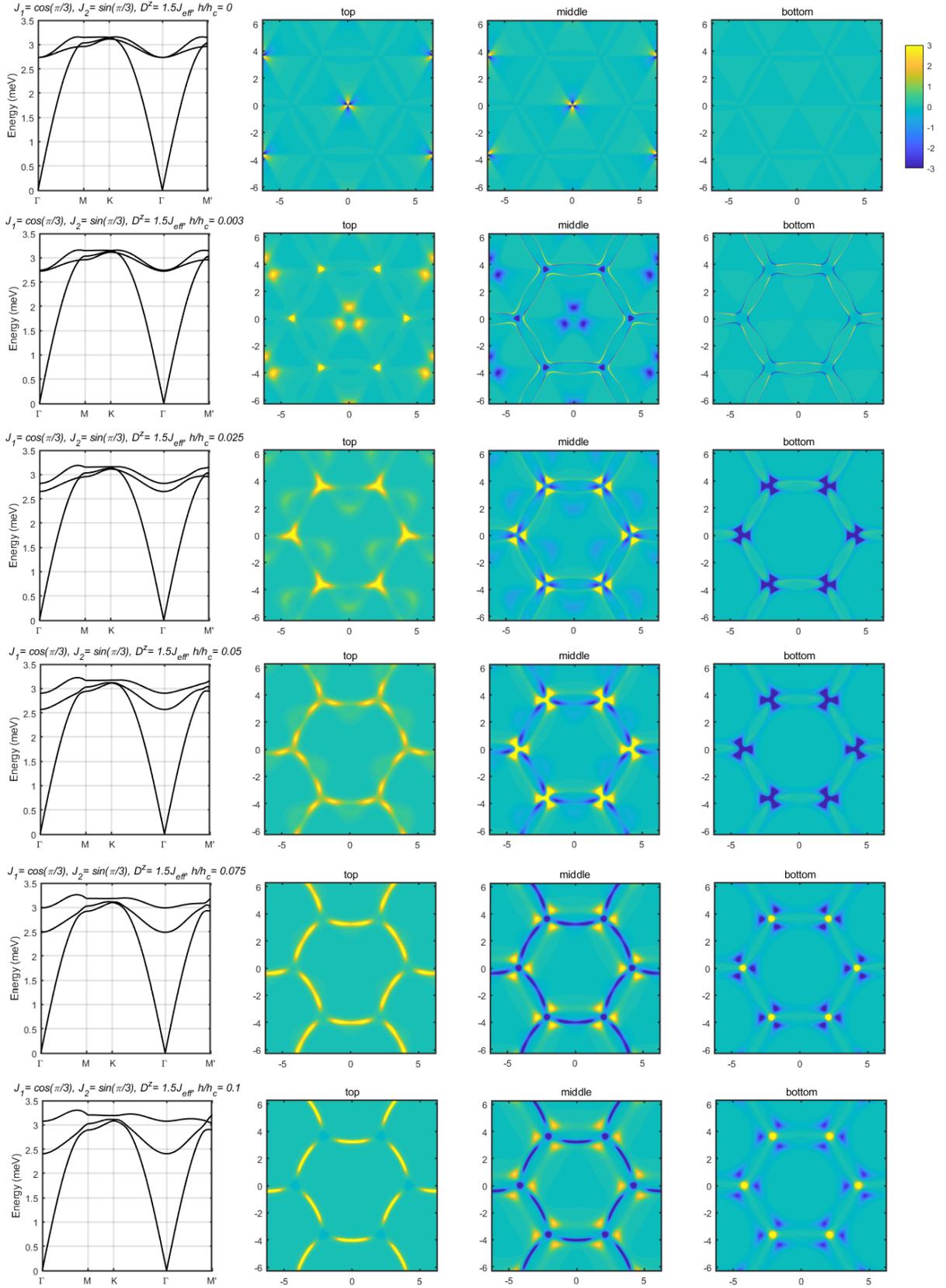

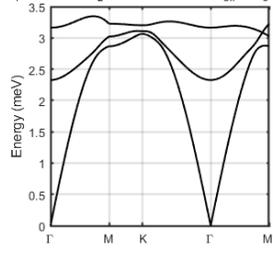 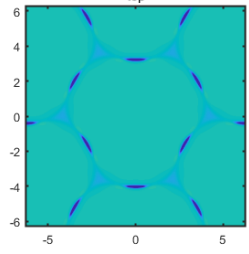 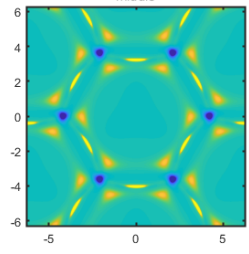 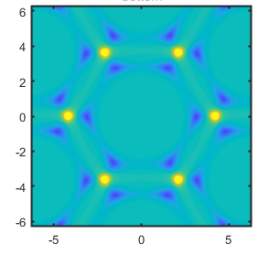
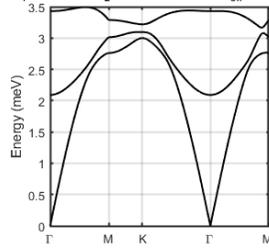 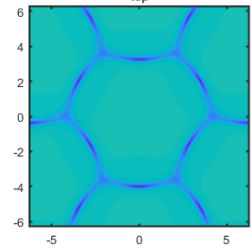 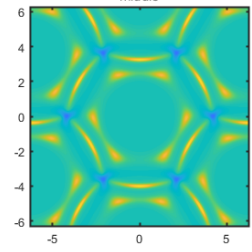 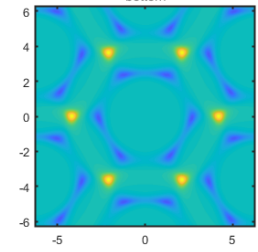
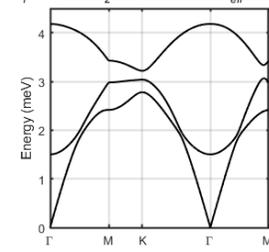 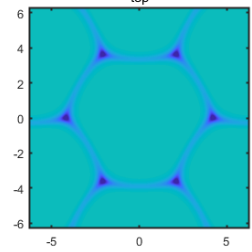 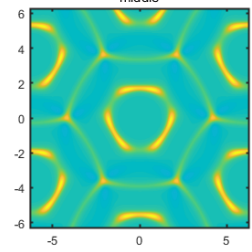 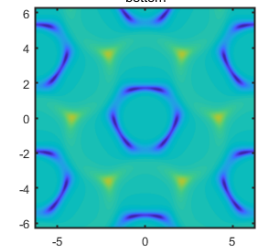
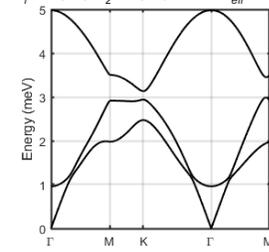 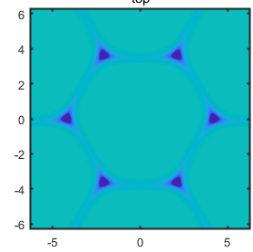 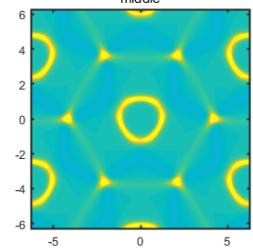 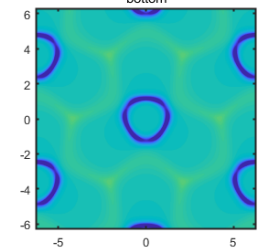
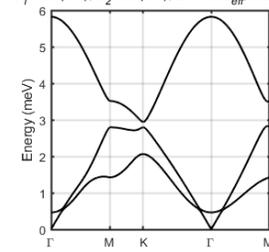 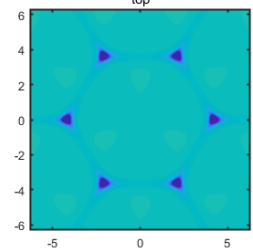 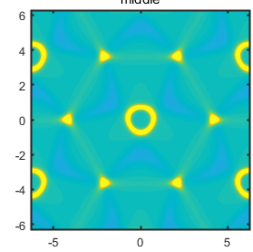 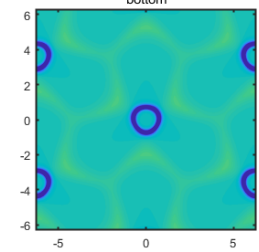
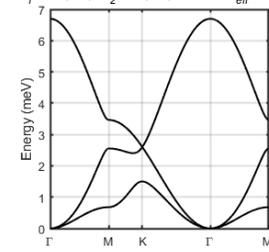 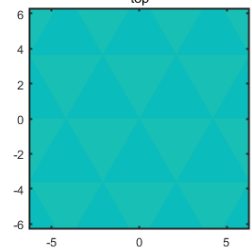 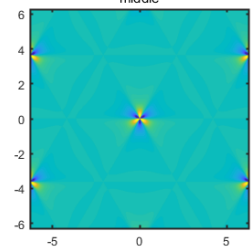 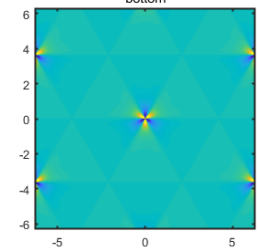

5

# 4 Fitting Results for $\kappa_{xy}$ near the band topology transition points

To confirm the singular behavior of $\kappa_{xy}$ at the topological band transition point, which is analytically derived in the main text, we fit the $\kappa_{xy}$ as a function of magnetic field $x$.

$$f(x) = A(x-x_0)\log(|x-x_0|) + \sum_{p=0}^{3} m_p (x-x_0)^p, \qquad (18)$$

where $x_0$ is the magnetic field at which the topological band transition occurs. The non-singular part is approximated by a cubic polynomial. We fit the result near topological band transition with parameters used in main text for YMnO$_3$: $J_1 = 2, J_2 = 2.4, D_z = 0.3\,\mathrm{meV}$ and $S=2$ at temperature $T=40\,\mathrm{K}$. Fitting parameters are shown in Tab. 1 and Fig. 2. The fittings are near exact and show that $(x-x_0)\log|x-x_0|$ represents well the singular part of $\kappa_{xy}$ as a function of the magnetic field. Note that even parity parts vanish at $x=0$ because of $\kappa_{xy}$ is odd under the magnetic field.

| $x_0$ | 0 | 5.84858 | 116.323 |
|---|---|---|---|
| $m_0$ | 0 | -1.66159 | 0.70561 |
| $m_1$ | -0.53804 | -0.0713343 | 0.0849511 |
| $m_2$ | 0 | 0.0118719 | 0.000176647 |
| $m_3$ | -0.496559 | -0.0025434 | $9.4649 \times 10^{-7}$ |
| $A$ | 0.490158 | -0.0117067 | -0.0112818 |

Table 1: Fitting parameters of $f(x)$ near band topology transition points.

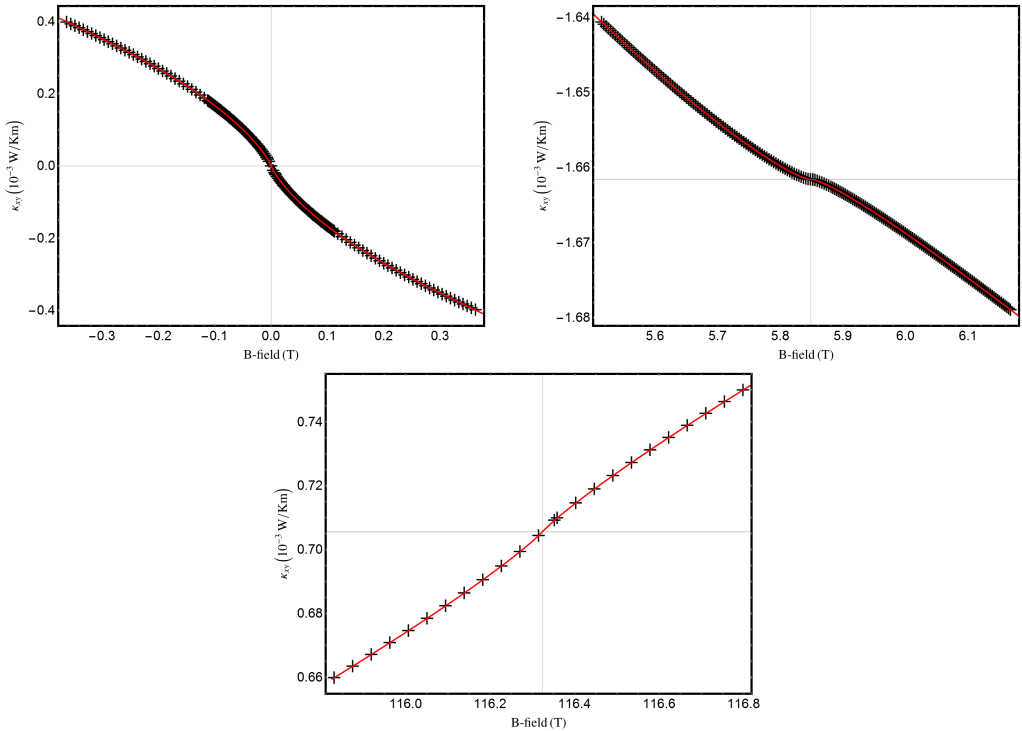

Figure 2: Black crosses are $\kappa_{xy}$ obtained by the numerical integration. Red lines are fitting functions which are near exact.



\end{document}